\documentclass[preprint,authoryear,12pt]{elsarticle}

\usepackage{pdflscape}
\usepackage{hyperref}
\usepackage{amsmath,amssymb,amsfonts}
\usepackage{geometry}
\usepackage{booktabs}
\usepackage{diagbox}
\usepackage{multirow}
\usepackage{makecell}
\usepackage{longtable}
\usepackage{subcaption}
\usepackage{graphicx}
\usepackage{comment}
\newtheorem{condition}{Condition}
\DeclareCaptionFormat{custom}
{
    \textbf{#1#2}\textit{\small #3}
}
\makeatletter
\def\ps@pprintTitle{%
\let\@oddhead\@empty
\let\@evenhead\@empty
\let\@oddfoot\@empty
\let\@evenfoot\@oddfoot
}
\makeatother

\begin{document}
\begin{frontmatter}
	
	
	
	\title{Sustainability Risks under Lotka-Volterra Dynamics}
	
	
	\author[1]{Yiren Wang}
    \author[1,2]{Tianhao Zhi}

\address[1]{Faculty of Science and Technology, Beijing Normal-Hong Kong Baptist University, Zhuhai, China}
\address[2]{Guangdong Provincial/Zhuhai Key Laboratory of Interdisciplinary Research and Application for Data Science, Beijing Normal-Hong Kong Baptist University, Zhuhai 519087, China}
	\cortext[1]{Corresponding author: Tianhao Zhi Email: tianhaozhi@bnbu.edu.cn}

	\begin{abstract}
The record-breaking heat in recent years, along with other extreme weather conditions worldwide has not only warned us about the devastating effects of global warming but also revived our interest in studying sustainability risks on a broader scale. In this paper, we propose a generalised model of sustainability risks characterising the economic-environmental-social nexus (EVS) based on a classic Lotka-Volterra framework. Compared to the \textit{World3} model proposed by \citet{M1972} in their landmark study \textit{``The Limits to Growth''}, our model has numerous advantages such as \textit{i)} better analytical tractability, \textit{ii)} more representative characterisation of economic development arising from innovation, and \textit{iii)} can be adopted in many potential applications of modelling sustainability risks from its sub-dynamics.
	\end{abstract}
	
	
\end{frontmatter}

\newpage
\section{Introduction}
The record-breaking heat in recent years, along with other extreme weather conditions worldwide, has not only warned us about the devastating effects of global warming but also revived our interest in studying sustainability risks on a broader scale. The Charter for the UCLA Sustainability Committee defines sustainability as: \emph{``the integration of \textbf{environmental health}, \textbf{social equity} and \textbf{economic vitality} to create thriving, healthy, diverse and resilient communities for this generation and generations to come} \citep{UCLASustainability2025}. The World Commission on Environment and Development defines sustainability as \emph{``meeting the needs of the present without compromising the ability of future generations to meet their own needs.''} \citep{WCED1987}. Subsequently, the UN advocates the Sustainable Development Goals (SDGs), which entail 17 specific goals that look at issues such as poverty, inequality, climate change, environmental degradation, peace and justice. The Conference of the Parties (COP) was initiated by the UNFCCC\footnote{United Nations Framework Convention on Climate Change}, established in 1992 at the Rio Earth Summit in Brazil, to address climate issues with a particular focus on emissions reduction. Since UNFCCC initiated the first COP\footnote{Conference of the Parties} meeting held in Berlin in 1995, there have already been 30 COP meetings held up to 2025. Over the past three decades, the COP meeting has evolved from its early function of establishing frameworks and protocols to its recent advocacy of specific actions, such as climate finance, carbon trading, and net-zero targeting.
\begin{figure}[h!]
	\centering
	\includegraphics[scale=0.3]{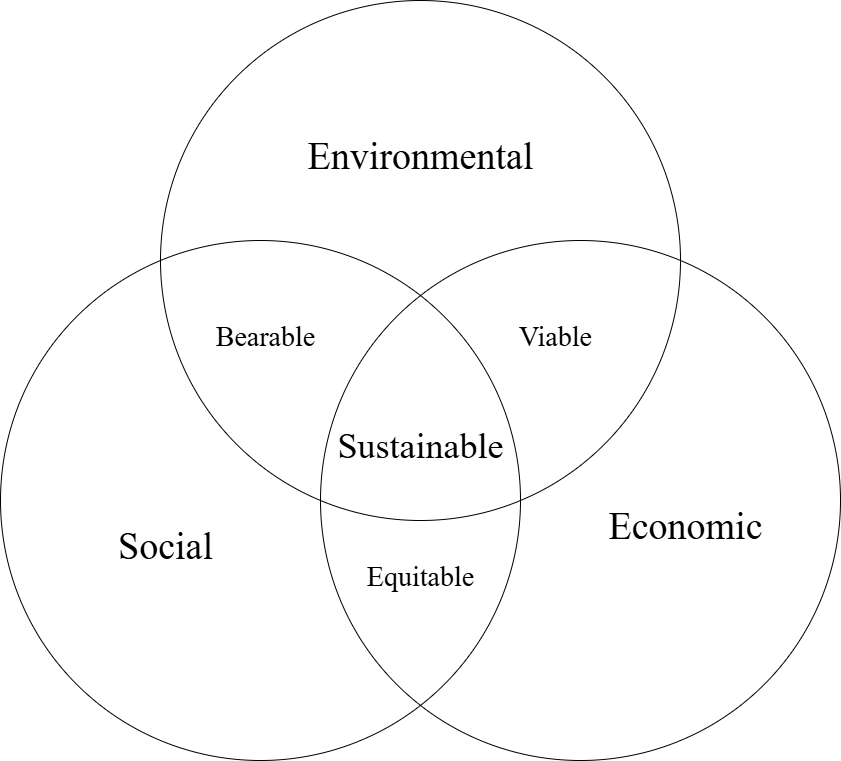}\\
	\caption{The Three Pillars of Sustainability \\ \textit{Source:} \citet{PMR2019}}
	\label{fig:sustainability}
\end{figure}

In finance, one of the most significant developments to address sustainability risks since the 1992 Rio Earth Summit is the introduction of the environmental, social, and governance (ESG) metric, which strives to establish a standard for disclosing sustainability-related information of public companies. Although the discussion of corporate social responsibility has a long history, ESG was not formally introduced and adopted until 2004, when the term ESG was coined in the UN-backed report \textit{Who Cares Wins} \citep{UN2004}. There is an increasing adoption of ESG reports by investors and an exponential growth of companies that start to provide ESG reporting, as well as agencies that offer ESG rating services. The dissemination of ESG has also led to an explosion of academic research on ESG in the field of corporate finance \citep{IG2019,BKR2022,CIS2014,DKL2015,AG2018,VCF2014,FBB2015,PFP2021,FKS2012}. By 25 October, 2025, a simple search with the term \emph{ESG} on \textit{sciencedirect} yields more than 17,000 results\footnote{See: \url{https://www.sciencedirect.com/search?qs=ESG} for the latest updates.}!

\begin{figure}[h!]
	\centering
	\includegraphics[scale=0.75]{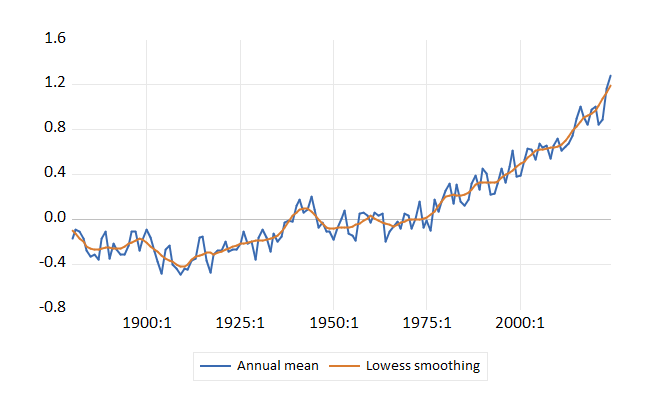}\\
	\caption{\citet{NASA2024}: Global Land-Ocean Temperature Index \\ \textit{Source:} NASA's Goddard Institute for Space Studies (GISS). Credit: NASA/GISS}
	\label{fig:GlobalTemp}
\end{figure}

However, there is little evidence that the climate issue has been alleviated over the last two decades since the introduction and adoption of ESG. According to Fig. \ref{fig:GlobalTemp}, there has been a steady rise in global temperature since 2000, which renders the years 2023 and 2024 the hottest two consecutive years in history \citep{Xie_etal_2025}. A recent CNBC news article reported that there are worries from the insurance industry that the world could \textit{soon become uninsurable} due to the rising global temperature \citep{CNBCUninsurable2025}. Both practitioners and academics raise questions regarding the efficacy of ESG in addressing climate issues. Some point to a lack of standardisation and rating divergence among the rating agencies, while others express their concerns about the scandals involving some high-profile public companies over recent years \citep{Johnson2025}. In academic research, most studies focus on trivial issues such as company performance, equity valuation, and investor trust, yet overlook the long-term efficacy of ESG in measuring long-term sustainability risks.

The dire situation of climate change today can be seen as a revelation of a more profound issue: while technological progress serves as a fundamental force of economic growth and human development, its unintended consequences can also lead to devastating sustainability risks that demand early intervention. Furthermore, it is not only the \textit{status quo}, but also the \textit{avant-garde} technologies that would pose significant sustainability risks with profound social and ecological consequences. While global warming is a culmination of decades-long consumption of fossil fuels since the advent of the internal combustion engine (ICE) over a hundred years ago, the disruptive technology of the digital era would potentially lead to an even greater spillover effect with a shortened time horizon of impact, both in positive and negative terms. The widespread adoption of AI may lead to a substantial rise in energy consumption and the displacement of traditional employment and social ethics \citep{Garcia2025}. The production of electronic gadgets may lead to severe water and soil pollution, as the production of such products requires a high consumption of lithium and rare earth materials that can have an adverse impact on the environment \citep{Duporte2022}. The adoption of electric vehicles may not reduce carbon emissions if the electricity is generated from coal instead of clean sources \citep{Zeng_etal_2021}. Sustainability risk is \textit{systemic}, whereby the failure of one part may lead to the collapse of the entire system \citep{SJ2024}. Therefore, it is vital to design a \textit{mission-oriented}\footnote{The term \textit{mission-oriented} innovation policy is coined by \citet{Mazzucato2013} in describing the critical role of the public sector in promoting innovation beyond the profit incentives of the private sector.} policy framework that not only encourages innovation but also guides R\&D spending to counterbalance the adverse effects of disruptive technology, with the ultimate goal of achieving a balanced growth trajectory that aligns with the UN’s Sustainable Development Goals.
%
%

\citet{KM2011} conduct a comprehensive survey of sustainability models. They base their analysis on the three pillars of sustainability (environment, economy, humanity) and categorise the sustainable development model into five groups: pictorial visualisation models, quantitative models, physical models, conceptual models, and standardising models. Within the last category of standardising model, the most well-known would be the landmark study \textit{``The Limits to Growth''} \citep{M1972}. The book is a commissioned work by the \textit{Club of Rome}, which had led to the creation of the \textit{World3} model by a team of MIT scientists: a large-scale computer model that simulates the long-term ramifications of unconstrained exponential economic growth by examining population, industrial output, food production, pollution, and consumption of non-renewable resources. The \textit{World3} model has been recalibrated and updated by \citet{M1992}, \citet{M2004}, and more recently by \citet{Herrington2021}. The original \textit{World3} model and its later variants generally assert that the global system will reach its limits within a foreseeable period in a \textit{business-as-usual (BAU)} scenario. Yet, humans have the capacity to avert the doomsday situation if preventive measures are taken.

Since the seminal work of \citet{M1972}, the \textit{World3} model has been the subject of considerable controversy and criticism. Numerous studies argue that the \textit{World3} model overlooks the forces of the market and innovation: the increasing scarcity of resources would raise their prices, forcing companies to find alternative resources in the short term and to innovate for other technological solutions in the long term \citep{Kaysen1972,Solow1973}. From a modelling perspective, several studies point out that the \textit{World3} model is sensitive to parameter changes and initial conditions \citep{Castro2012}. Indeed, the \textit{World3} model is a high-order, non-linear dynamic system with 12 state variables and internal delay functions with complex and intertwined feedback loops, which renders it theoretically challenging to derive its closed-form analytical solution \citep{PyWorld3}. It is widely recognised that the predictive horizon of complex models is constrained by their degree of complexity, particularly for the case of a chaotic system \citep{Lorenz1963}. It is questionable whether the very long-term trend projection derived from such a complex model is valid over the very long time horizon. On the contrary, a dynamic model with a more parsimonious structure may offer better predictive power over the short to medium term, which is vital for setting the stage for policy debates and guiding regulatory measures over a shorter time horizon. Furthermore, the market economy is a dynamic process constantly fueled by disruptive innovation and incessant entrepreneurial pursuits. Hence, the model should capture the sustainability risk resulting from the \textit{creative destruction} process in Schumpeterian terms \citep{Schumpeter1934}. Despite its controversy, the \textit{World3} model has paved the way for researchers later on to study sustainability risks from a systems dynamics perspective.

One possible candidate model that can address this issue is the classical Lotka-Volterra (LV) model (also known as the predator-prey model), which examines the dynamic competitive and cooperative relationships among the state variables as predators or prey. Since the seminal work of \citet{Lotka1920} and \citet{Volterra1926}, many studies have extended the classic pairwise 2D LV system to include richer mathematical structures \citep{Vandermeer1990,GOUZE1993}. The most recognised application of the LV model is in biology, yet it has found a wide range of applications in environmental science and even economics \citep{Goodwin1982,Keen2013,Sun2023}. Despite its simple mathematical structure, the LV model can be applied to a wide range of problems and is highly relevant in modelling the sustainability of the ecological, social and economic systems.

In this paper, we propose a generalised model of sustainability risks characterising the economic-environmental-social nexus (EVS). Our model has a simplistic character in its mathematical structure. Yet it has a high degree of flexibility that can be applied to understand sustainability issues in a generalised and unified framework. Compared to the \textit{World3} model, our model has numerous advantages. \textit{First}, our model has better analytical tractability, as the Lotka-Volterra model and its variants are well-studied in the fields of applied mathematics and ecology. \textit{Second}, our model can better capture innovation-led economic development, addressing the critique of the \textit{World3} model that neglects the force of innovation. \textit{Third}, our model can better characterise sustainability risks in a generalised and flexible manner, as the model can be sub-divided in many subdynamics that tackle specific sustainability risks of each sub-sector.

The rest of the paper is organised as follows: Section \ref{MSU} sets up the model in its generalised form, with a discussion regarding the measurement and scope of the three state variables. Section \ref{subd} looks at the three subdynamics relating to economic-environmental (EV), economic-social (ES), and environmental-social (VS) nexus. Section \ref{3D} examines the full 3D EVS system. Section \ref{nd} proposes an extension of an $N$-dimensional EVS system for the interests of future research. Section \ref{Conclusion} concludes.
\section{Model Set-up}
\label{MSU}
We use the letters \textit{E}, \textit{V}, \textit{S} to represent the economy, the environment, and the society\footnote{\citet{KM2011} use the letters \textit{E}, \textit{H}, \textit{N} to denote economy, humanity, and environment. In this study, we use \textit{V} instead of \textit{N} to eliminate confusion associated with the \textit{N}-dimensionality that we refer to in the latter section. We use \textit{S} (social) to replace \textit{H} to generalise social development issues, in which humanity is undeniably a vital component.}. The three variables represent a generalised depiction of the economic, social, and environmental states, which can be decomposed into interconnected contributing factors denoted as $E_i$, $V_i$, and $S_i$. The three state variables are set in a Lotka-Volterra framework where they would either be cooperative or competitive with each other:
\begin{eqnarray}
\dot{x}_i &=& x_i (r_i + \sum^{n = 3}_{j=1} a_{ij} x_j), i = 1, 2, 3,\\
x_1 &=& E,\\
x_2 &=& V,\\
x_3 &=& S,
\label{eq1}
\end{eqnarray}
\noindent where $r_i$ is the intrinsic growth rate of each state variable. $a_{ij}$ describes the quadratic interactions between the three state variables. $x_i$ contains the three state variables: \textit{E}, \textit{V} and \textit{S}.

\subsection{The definition of the three state variables}
This section discusses the measurement and the components of the three state variables, which is by no means a simple task. The commonly adopted measures of economic, social, and environmental indicators often have their limitations, and sometimes they overlap with one another in certain sub-dimensions. It is necessary to provide a taxonomy to facilitate further analysis regarding their competitive and cooperative relationships in the Lotka-Volterra setting.

\subsubsection{Measuring economic development}
The letter \textit{E} represents not only the economy, but also entrepreneurship. From a Schumpetarian perspective, such measures should be able to characterise economic development contributed by entrepreneurship and innovation. It also should reflect the economic impacts arising from the continuous accumulation of knowledge, either proprietary knowledge in the form of various kinds of IP such as patents or copyrights, or common, open source knowledge that has real impacts on productivity and economic growth. Both the existing knowledge (the \textit{status quo}) and the emerging new knowledge (the \textit{avant-garde}) would contribute to economic development, with varying degrees of impact over their lifetime. One may be tempted to adopt simple macroeconomic measures such as GDP, which is defined as \textit{``the total market value of all final goods and services produced within a country in a given period of time''} \citep{BHM2009}. We argue that it is inappropriate to take this oversimplistic approach for two reasons. \textit{First}, GDP represents a \textit{flow}, rather than a \textit{stock}, of a given period, which renders it incomparable with other environmental and social indicators that are \textit{stocks} in nature. A deeper issue concerns the distinction between economic \textit{growth} and economic \textit{development}. While growth can be measured in standardised and quantitative terms, economic development involves more profound structural changes beyond quantitative measures. J.A. Schumpeter's \textit{Theory of Economic Development} depicts the vital role of entrepreneurs in the process of economic development, as they constantly seek profit by innovation, resulting in the creative destruction\footnote{The term Creative Destruction was not coined until Schumpeter's later works \textit{Capitalism, Socialism, and Democracy} \citep{Schumpeter1942}.} and structural changes that serves as the ultimate engine of economic development \citep{Schumpeter1934}. It is therefore necessary to integrate innovation-oriented measures such as the Global Innovation Index\footnote{The Global Innovation Index (GII), launched in 2007 by the World Intellectual Property Organisation (WIPO), provides a comprehensive framework in measuring innovation on a worldwide basis \citep{Dutta2007,WIPO2024}.} (GII) as a measurement of economic development. The innovation-led productivity growth, however, is not a guarantee for sustainable development. The \textit{Jevons Paradox} is the observation that improvements in resource efficiency can lead to a rising demand of that resource, thus posing sustainability risks \citep{Jevons1865}. Therefore, it is equally important to measure social development and environmental quality to gain a comprehensive understanding of sustainability risks arising from innovation, which will be further elaborated in the following sections.


\subsubsection{Measuring social development}
Social development encompasses issues concerning the well-being of humanity. The Centre for Sustainable Systems (CSS) at the University of Michigan identifies eight key social development issues on an universal basis, including \textit{i)} population, \textit{ii)} standard of living, \textit{iii)} food, \textit{iv)} water and sanitation, \textit{v)} healthcare and disease, \textit{vi)} education and employment, \textit{vii)} environment, \textit{viii)} global initiatives \citep{CSS2024}. The definition and measurement of social development, however, often intertwine with that of economic development. Compared with economic development issues that are primarily and universally profit-driven and productivity-oriented, social development issues are much more multidimensional, often requiring culturally specific measures and strong public policy support in areas where the \textit{laissez-faire} market force does not deliver.

Yet the CSS measure of social development overlooks inequality as a key issue of social development. The earliest and perhaps most well-known inequality measure is the Gini index, proposed by \citet{Gini1912}, where inequality is measured on a scale from $0$ (perfect equality) to $1$ (perfect inequality). More recently, \citet{Aspalter2023} systematically examines the causes of \textit{super inequality} from his theory of \textit{Z-efficiency} that addresses a broader scope of social, behavioural, and institutional inefficiency. The $Z$-efficiency theory is extended from Leibenstein's $X$-efficiency theory of \citet{Leibenstein1966}, and it encompasses a set of multi-dimensional social efficiency measures, including $M$-efficiency (removal of managerial barriers), $N$-efficiency (removal of environmental and physical barriers), $Y$-efficiency (removal of personal-level barriers), and $Z$-efficiency (removal of social and cultural barriers). The $Z$-efficiency theory can serve as an important measure of social development in our context.
%
%

Another aspect that the \citet{CSS2024} overlooks is the cultural dimension of social development, which is difficult to measure in quantitative terms, yet is deeply connected with economic and environmental dimensions. The abundance of cultural heritages in a mature society contributes significantly to a better environment, as well as better economic opportunities through sustainable fishing, farming, and other activities. The positive link between ICH and sustainable development is well-documented in numerous studies \citep{Meissner2021,BS2023,Bortolotto2025}. More efforts are needed to establish a quantitative framework for gauging cultural heritage as a culturally specific dimension of social development in future research.

\subsubsection{Measuring environmental quality}
\begin{figure}[h!]
	\centering
	\includegraphics[scale=0.25]{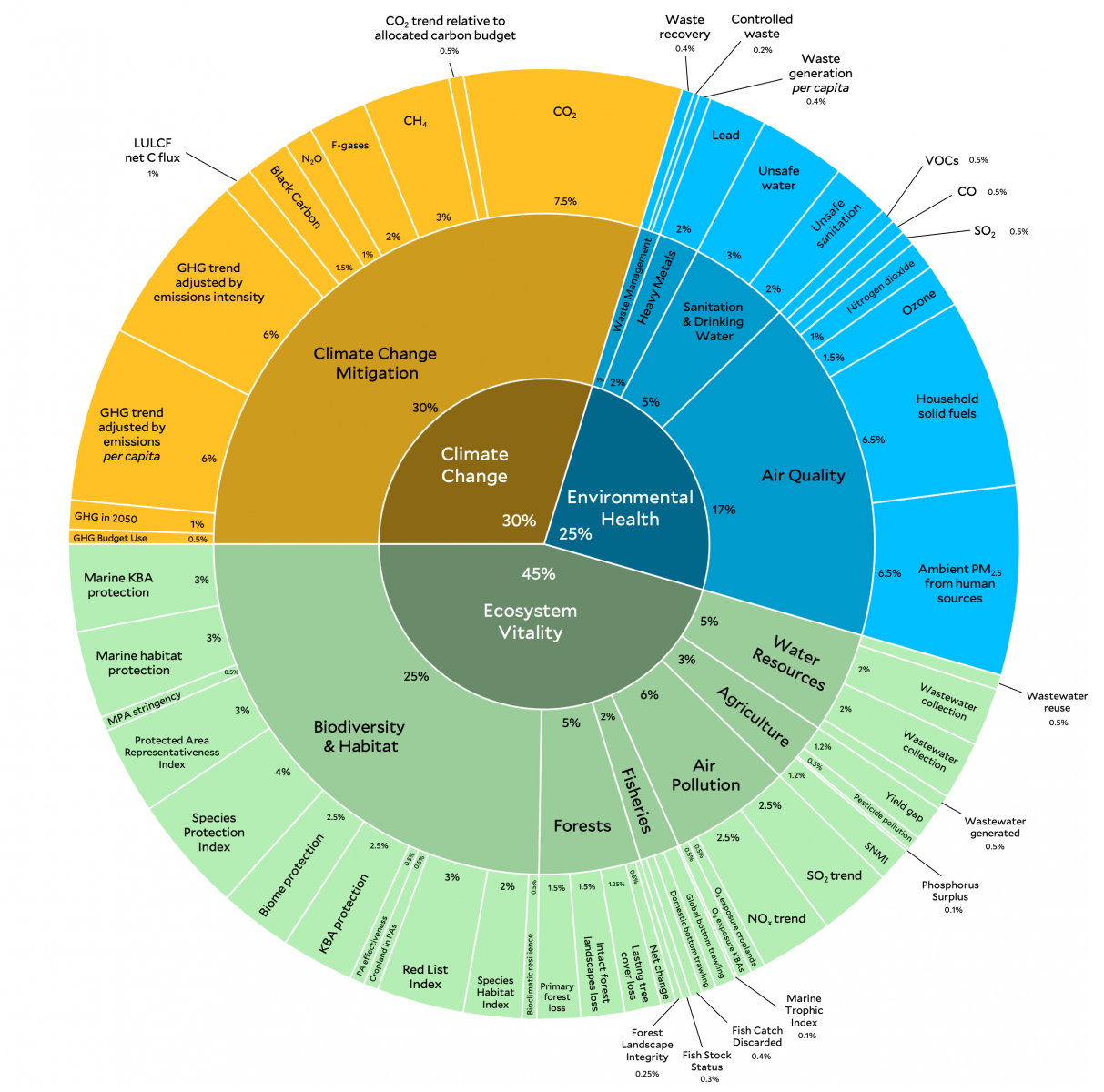}\\
	\caption{\citet{EPI2024}: The 2024 Weighting Pie of the Environmental Performance Index (EPI) \\ \textit{Source:} The Yale Centre for Environmental Law and Policy \\ Figure adopted from the following URL: \url{https://epi.yale.edu/}, with the permission from the authors}
	\label{fig:EPI}
\end{figure}

Climate issues resulting from global warming have received the most attention from social media and academic studies. It is a culmination of decades of fossil fuel consumption since the early industrial revolution in the 18th century. However, there are many other potential environmental threats arising from the era of digitalisation beyond fossil fuel consumption that deserve further investigation so that preventive measures can be taken to avoid ecological crises. The Yale Centre for Environmental Law and Policy proposes the Environmental Performance Index (EPI), which compiles a comprehensive global measurement of environmental performance across three broad categories: climate change performance, environmental health, and ecosystem resilience. Within these three categories, there are 58 performance indicators across 11 issue categories, referencing 180 countries, as illustrated in Fig. \ref{fig:EPI} \citep{EPI2024}. It can serve as a comprehensive measure of environmental quality in our context.

\section{The EV, ES, VS subdynamics}
\label{subd}
We begin with a general discussion of the three subdynamics with a particular focus on the signs of the intrinsic growth parameters $r_x$ and the interactive parameters $a_{ij}$. The discussion here is not absolute, but a stylised depiction of how the baseline model may look like. As for the growth parameters, we argue that only $r_V$ has a positive sign, while $r_E$ and $r_S$ are assumed to be negative for the simple reason that neither the economy nor the society would survive in the absence of a well-functioning ecological system. The interactive parameters deserve some further discussion in the following subsections.
\subsection{The EV (economic-environmental) subdynamics with the viable scenario}

The EV subdynamics can be written as:
\begin{eqnarray}
    \dot{E} &=& E(r_E +a_{E}V), \\
    \label{eqn_EVsub_E}
    \dot{V} &=& V(r_V +a_{V}E).
\end{eqnarray}

The E-V nexus is typically characterised by a standard predator-prey relationship ($r_E < 0$, $a_E > 0$, $r_V > 0$, $a_V < 0$) in most existing studies \citep{EGO2020,Sun2023,OAY2025}. The resource constraint on economic growth is also the backbone of the \textit{World3} model. We denote $r_E$ as the rate of economic contraction, as opposed to a standard assumption in neoclassical growth theory that an economy can grow exponentially based on either an exogenous or an endogenous assumption, for the simple reason that without an environmental endowment and in the absence of social development, the primitive economy will have no other options but to decay. In such a highly stylised hunter-gather society, the economic development rate is determined by the environmental utilisation ratio $a_E$, which represents the proportion of environmental stock being utilised to fuel economic development $\dot{E}/E$. On the other hand, we assume that $r_V > 0$, implying that the environmental factor $V$ can grow exponentially without the presence of human activities. This is a highly simplified assumption, as the ecological system would also reach its limits as depicted by a classical logistic growth model. Here we are making this assumption only for mathematical convenience in a Lotka-Volterra setting. Compared with $a_E$ that measures how environmental stock contributes to economic growth (the \textit{from} effect), $a_V$ measures how economic development leads to environmental deterioration (the \textit{to} effect).

As a standard Lotka-Volterra model, there exists two fixed points: one zero ($E=0$ and $V=0$) and the other non-zero (we may call this \textit{viable}, based on the aforementioned Venn diagram of sustainability). The non-zero fixed point can be derived as follows:
\begin{eqnarray}
    0 &=& r_E +a_{E}V, \\
    0 &=& r_V +a_{V}E, \\
    \Rightarrow V^\star &=& -r_E / a_E, \\
    E^\star &=& -r_V / a_V.
\end{eqnarray}

The Jacobian of the EV subdynamics is calculated as:
\begin{eqnarray}
	J&=&
	\begin{bmatrix}
		r_E+a_EV & a_E E,
		\\
	a_V V & r_V+a_V E,
	\end{bmatrix},
\end{eqnarray}

\noindent where the trace and determinant at the \textit{viable} fixed point are calculated as
\begin{eqnarray}
   Tr(J) &=& r_E+a_E V^\star+r_V+a_V E^\star = 0,\\
   Det(J) &=& (r_E+a_E V^\star)(r_V+a_V E^\star)-a_E E^\star a_V V^\star = - r_E r_V > 0.
\end{eqnarray}

This replicates the results of a standard 2D Lotka-Volterra model. It implies the existence of closed orbits in the LV system according to the Poincar\'e diagram \citep{Strogatz2018}, which forms the backbone of the cyclical behaviour of the 3D model that we will discuss later in Section \ref{3D}.


\subsection{The ES (economic-social) subdynamics with the equitable scenario}
The ES subdynamics is written as:
\begin{eqnarray}
\label{eqn_ESsub}
    \dot{E} &=& E(r_E +a_{E}S), \\
    \label{eqn_ESsub_E}
    \dot{S} &=& S(r_S +a_{S}E).
\end{eqnarray}

The mathematical structure follows a canonical form compared with the EV subdynamics, except that the parameters may take different signs. The economic-social nexus was metaphorically described as \textit{the base and the superstructure} by  K. Marx, as he wrote ``\textit{the changes in the economic foundation lead, sooner or later, to the transformation of the whole immense superstructure}'' \citep{Marx1859}. While economic development can undoubtedly support social development, it also gives rise to the problem of income inequality, which remains a significant constraint that hinders social development, from the time of Marx to the present. In the modern-day era, inequality is often associated with the overexpansion of the financial service industry and the concept of financialisation: a process \textit{``whereby financial markets, financial institutions, and financial elites gain greater influence over economic policy and economic outcomes''} \citep{Palley2007}. The widening income gap between the financial sector and the non-financial sector is documented in T. Piketty's monumental study, \textit{Capital in the Twenty-First Century} \citep{Piketty2014}. Beyond the financial service industry, the confinement of technological innovation to a small handful of tech elitists is another fundamental source of inequality of our time, characterised by \textit{pseudo-vanguardism}, \textit{hyper-insularity}, and \textit{precarious employment}, as vividly depicted in \textit{ The Knowledge Economy} written by R. M. Unger \citep{Unger2019}. 

We set $r_E < 0$, $r_S < 0$, $a_E > 0$, while the sign of $a_S$ remains uncertain, as Smith, Marx, Keynes and Schumpeter may disagree with each other.

%
\subsection{The VS (environmental-social) subdynamics with the bearable scenario}
The VS subdynamics is written as:
\begin{eqnarray}
    \dot{V} &=& V(r_V +a_{V}S), \\
    \label{eqn_ESsub_E}
    \dot{S} &=& S(r_S +a_{S}V).
\end{eqnarray}

The ecological system is the foundation of all human activities, without which neither the economic nor the social system could exist. The interaction between social and environmental systems can be characterised as mutually supportive \citep{ODA2022,Barron_etal_2023}. A rich environmental endowment can foster social development, particularly at an early stage. For example, the Tigris-Euphrates river system had fostered the early Mesopotamian civilisation, which serves as a representation of many other similar historical references. Better social development can, in turn, support environmental preservation through better education, social innovation, and full utilisation of green infrastructure and human capital. Therefore, we expect that $r_V > 0$, $r_S < 0$ - reflecting the self-sufficiency of the ecological system and the dependence of society on the ecological system; $a_V > 0$ and $a_S > 0$ - reflecting the mutually supportive role between \textit{V} and \textit{S}.
\subsection{The merged full model and the feedback loops}
We summarise the feedback loops of the EV, ES and VS subdynamics in Fig. \ref{fig:EVS_feedback} below. The mathematical property of the full 3D EVS system will be discussed in Section \ref{3D}.
\begin{figure}[h!]
	\centering
	\includegraphics[scale=0.4]{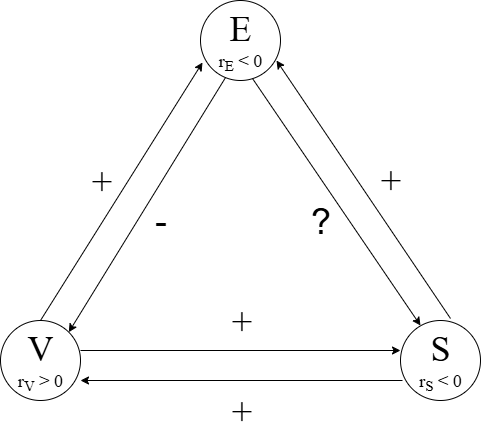}\\
	\caption{A stylised feedback loop between E, V, and S}
	\label{fig:EVS_feedback}
\end{figure}

\section{The full 3D EVS system}
\label{3D}
\subsection{The baseline model}
We merge the EV, ES, VS subdynamics to form the baseline 3D EVS model in Eq. \ref{eqn_3D_E} - \ref{eqn_3D_S}. 
\begin{eqnarray}
\frac{dE}{dt} &=&E(r_E + a_{12}V + a_{13}S),
\label{eqn_3D_E}
\\
\frac{dV}{dt} &=&V(r_V + a_{21}E + a_{23}S),
\label{eqn_3D_V}
\\
\frac{dS}{dt} &=&S(r_S + a_{31}E + a_{32}V),
\label{eqn_3D_S}
\end{eqnarray}

\noindent where the three intrinsic growth rate $r_E<0$, $r_V>0$, $r_S<0$. The interactive parameter $a_{ij}$ ($i,j=1,2,3$) replace $a_E$,  $a_V$, and  $a_S$. Based on previous section, we set $a_{12}>0$, $a_{13}>0$, $a_{21}<0$, $a_{23}>0$, $a_{32}>0$, while the sign of $a_{31}$ would either be positive or negative. 

To further examine the dimension of economic development, Eq. \ref{eqn_3D_E} can be rewritten as:
\begin{eqnarray}
E \% \approx \frac{\dot{E}}{E} = r_E + a_{12}V + a_{13}S.
\label{eqn_3D_E2}
\end{eqnarray}

Eq. \ref{eqn_3D_E2} encompasses two types of economic development: \textit{resource-driven} development and \textit{socially supported} development. It is broadly in line with the classic growth theory \citep{Solow1956,JM1997}, except that $r_E<0$ as discussed previously. Resource-driven development is the most primitive form, while socially supported development is an advanced form of economic development. High social development entails a well-developed social infrastructure and the availability of human capital, which supports economic development through innovation and creativity.

%
%
%
%


\subsection{The sustainability and resilience conditions of the EVS system}
\begin{figure}[h!]
	\centering
	\includegraphics[scale=0.3]{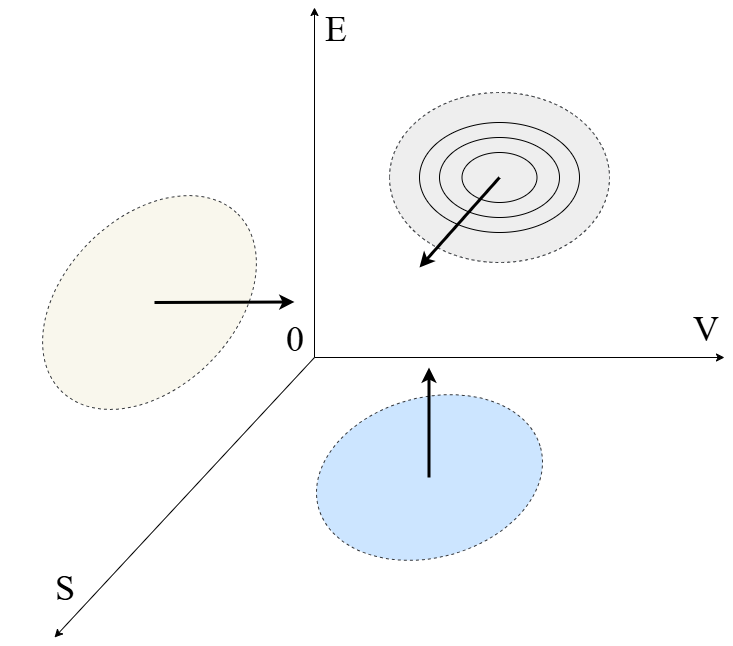}\\
	\caption{An illustration of the sustainability and resilience conditions}
	\label{fig:sustainability}
\end{figure}
We take a further step to investigate how sustainability can be modelled in the EVS system. Recall that WCED\footnote{The World Commission on Environment and Development, established by the UN and led by former Norwegian Prime Minister Gro Brundtland} defines sustainable development as \textit{meeting the needs of the present without compromising the ability of future generations to meet their own needs} \citep{WCED1987}. We may translate this definition as the three state variables in the EVS system that coexist with, and are interdependent upon each other, so that none will become suppressed to extinction due to the overexpansion of the other(s). Mathematically, this can be formally defined as:

\noindent \textbf{Definition 1 (Sustainability):} \textit{An EVS system is sustainable when}
\[\exists \text{ } T_1 \text{ such that when } T > T_1, \ E> 0, V> 0, S > 0\text{.}\]

In the current academic literature, the terms \textit{sustainability} and \textit{resilience} are often used interchangeably. However, the two concepts may have substantial differences in the practice of sustainable development and risk management. \citet{RNK2019} conduct a meta-analysis on the definitions of, and relationship between sustainability and resilience. They point out that sustainability can simply mean \textit{``to maintain the status quo and to not disappear''}, while resilience refers to \textit{``the responses to shocks, surprises, unforeseen or hazardous disturbances''}, thereby \textit{``having the capacity to persist in the face of change, to continue to develop with ever-changing environments''}. In the context of the EVS system, resilience implies that the system can recover on its own in the event of near-extinction or near-destruction due to external shocks. Mathematically, we may define resilience as:

\noindent \textbf{Definition 2 (Resilience):} \textit{An EVS system is \textit{resilient} when}
\begin{eqnarray}
\mathbf{X}(0) = (E(0), V(0), S(0)) &\in& \mathbb{R}^3_{+},
\\
\frac{dE}{dt} \bigg|_{E \to 0} &>& 0,
\\
\frac{dV}{dt} \bigg|_{V \to 0} &>& 0,
\\
\frac{dS}{dt} \bigg|_{S \to 0} &>& 0.
\end{eqnarray}

It is difficult to derive a general condition for global sustainability under the EVS system specified by Eq. \ref{eqn_3D_E} - \ref{eqn_3D_S}, since the closed orbit from the EV subdynamics does not have an exact functional expression. However, it is possible to derive the sustainability condition at the fixed point of the subdynamics to determine whether it exhibits the resilience condition or not, following \citet{Vandermeer1990}. For example, in the case of an EV subsystem, we may set $S=0$ to find the fixed point at the EV subsystem $(E^\star,V^\star)$ and set $\frac{dS}{dt} > 0$ to further derive its resilience condition. We may interpret this equation in a hypothetical situation of near-complete social destruction. With the injection of a tiny institutional aid, the sign of $\frac{dS}{dt}$ would determine whether the social system will recover on its own or not. The other two state variables can be examined in the same manner, as illustrated in Fig. \ref{fig:sustainability}. Formally, we define that:

\noindent \textbf{Proposition 1 (Fixed Point Resilience):} \textit{The fixed point $(E^{\star}, V^{\star})$ of the EV sub-system defined by Eq. \ref{eqn_3D_E} - \ref{eqn_3D_S} is \textit{resilient} on the social dimension when the following condition set by Eq. \ref{eq:const1} is fulfilled:}

\begin{equation}
\left\{
\begin{aligned}
a_{21} < \dfrac{a_{31}a_{12}r_V}{a_{12}r_S - a_{32}r_E} &\  \text{if }\ a_{12}r_S > a_{32}r_E,  \\\\
a_{21} > \dfrac{a_{31}a_{12}r_V}{a_{12}r_S - a_{32}r_E} &\  \text{if }\ a_{12}r_S < a_{32}r_E.  
\end{aligned}
\right.
\label{eq:const1}
\end{equation}



\noindent \textit{Proof:} we first set $S = 0$, rendering the system an EV subdynamics:
\begin{equation}
\left\{
\begin{aligned}
\frac{dV}{dt} &= V(r_V + a_{21}E) = 0, \\
\frac{dE}{dt} &= E(r_E + a_{12}V) = 0,
\end{aligned}
\right.
\end{equation}

\noindent which yields

\begin{equation}
\left\{
\begin{aligned}
E^* &= -\frac{r_V}{a_{21}},
\\
V^* &= -\frac{r_{E}}{a_{12}}.
\end{aligned}
\right.
\end{equation}

We can see that $E^{\star} > 0$ and $V^{\star} > 0$, since $r_E < 0$, $a_{12} > 0$, $r_V > 0$, and $a_{21} < 0$. Furthermore, the social resilience condition implies that:
\begin{equation}
\frac{dS}{dt} = S(r_S + a_{31} E + a_{32} V) > 0.
\end{equation}

Hence,
\begin{equation}
r_S > \frac{a_{31} r_V}{a_{21}} + \frac{a_{32} r_E}{a_{12}} = \frac{a_{31} a_{12} r_V + a_{32} a_{21} r_E}{a_{21}a_{12}}.  
\end{equation}

Since we assume that $a_{21} < 0, a_{12} > 0, a_{32} > 0, r_S<0$, and $\ r_E < 0$, it implies that:
\begin{eqnarray}
a_{21} (a_{12} r_S - a_{32} r_E) &<& a_{31} a_{12} r_V,
\end{eqnarray}

\noindent which leads to:

\begin{equation}
\left\{
\begin{aligned}
a_{21} < \dfrac{a_{31}a_{12}r_V}{a_{12}r_S - a_{32}r_E} &\  \text{if }\ a_{12}r_S > a_{32}r_E,  \\\\
a_{21} > \dfrac{a_{31}a_{12}r_V}{a_{12}r_S - a_{32}r_E} &\  \text{if }\ a_{12}r_S < a_{32}r_E.  
\end{aligned}
\right.
 \label{eq:VE}
\end{equation}

Similarly, it is also possible to derive the resilience conditions for the fixed points of the ES and VS system. However, the fixed points may be unattainable due to the constraints we set on the signs of the $r_{k}$ and $a_{ij}$ parameters, rendering some values of the fixed points negative. We put the mathematical derivation of the fixed points for the ES and VS system, along with its resilience conditions in \ref{app_1}.

We augment Definition 1, Definition 2 and Proposition 1 with a critical economic interpretation: the trajectory of motion under a particular parameter set represents a \textit{bottom-up} outcome. In contrast, the parameter set itself represents a \textit{top-down} precondition that can be altered through public intervention. Sustainability risks may arise not only from existing technology (as epitomised by the consumption of fuels), but also from the unpredictable, disruptive innovation (the \textit{Schumpeterian} part). The emergence of sustainability risks is characterised by the violation of the aforementioned sustainability condition within a particular parameter set. The public sector, however, can alter the parameter set through various forms of intervention, such as fiscal stimulus, environmental regulation, and mission-oriented innovation policies to ensure that the parameter set remains within the sustainable regime (the \textit{Keynesian} part). In other words, certain parameter sets would determine whether the \textit{laissez-faire} market force would fail to deliver the desired sustainable development outcome, and whether public intervention is needed to ensure sustainability. This interpretation aligns well with the \textit{Keynesian-Schumpeter synthesis} discussed by \citet{Dosi2010}, as well as the concept of \textit{the entrepreneurial state} and \textit{mission-oriented innovation policies} proposed by \citet{Mazzucato2013}.

\subsection{Numerical simulations}
We run three sets of simulations to visualise the 3D baseline EVS model. The simulation results are provided in \ref{appendix_NS}. The parameters are set as follows: $r_E = -0.1$, $r_V = 0.1$, $r_S=-0.05$, $a_{12} = 0.7$, $a_{13}= 0.1$, $a_{21} = -0.3$, $a_{23} = 0.1$, $a_{31} = \pm 0.1$, $a_{32} = 0.1$. The initial condition is set as $\mathbf{X}(0) = (E(0), V(0), S(0)) = (0.1, 0.1, 0.1)$. The first two sets of simulation in Fig. \ref{fig:EVS}-\ref{fig:EVSS} compares the scenario when $a_{31} = 0.1$ and $a_{31} = - 0.1$. The first set of simulations in Fig. \ref{fig:EVS} illustrates a scenario where social and economic development form a mutually cooperative relationship, leading to the harmonic co-existence between the economic, social and environmental development, despite the existence of perpetual cyclical behaviour induced by the EV subdynamics. The second set of simulations in Fig. \ref{fig:EVSS} draws the simulation when $a_{31}$ is flipped to a negative value of $-0.1$, \textit{ceteris paribus}. It implies that economic development hinders social development, rendering the system a so-called \textit{viable} scenario where economic and environmental development were preserved at the cost of social development.

The third simulation set in Fig. \ref{fig:EVS_bif} examines the sensitivity of the system when the parameter $a_{12}$ varies from $0.6$ to $1.5$. The most noticeable difference lies in the upper-right simulation associated with the sensitivity of $V$ when $a_{12}$ increases: in the first scenario of the sub-figure \ref{fig:EVS1_bif}, where there is a mutually cooperative relationship between $E$ and $S$, $V$ exhibits a hump-shaped form. This indicates that in such a scenario, there exists a higher tolerance of environmental predation, since the added benefits to social dimension from economic development can also indirectly contribute to better environmental outcome, compared with the inverse relationship in the sub-figure \ref{fig:EVS2_bif} that depicts a direct inverse relationship between environment $V$ and environmental predation $a_{12}$.

%
  
%
%
\section{A generalised extension to an $N$-dimensional EVS model}
\label{nd}
The previous section provides a thorough analysis of the baseline 3D EVS system. The state variables take an aggregated form, and the interactive signs ($a_{ij}$) are set in a highly stylised manner in the baseline model. To take one further step, each state variable consists of many contributing factors, and these contributing factors may be \textit{inter}-connected across the three dimensions or \textit{intra}-connected within a particular dimension - for example, the \textit{creative destruction} process can be manifested as a competitive LV relationship within the $E$ subdimension between the new technology and the old technology. At the same time, the development of quantum computing and AI can be characterised by a cooperative LV relationship. We set up an $N$-dimensional Lotka-Volterra model that consists of $N = n_1 + n_2 + n_3$ variables, in which there are $n_1$ economic factors, $n_2$ environmental factors, and $n_3$ social factors, expressed in Eq. \ref{eq:nd1} - \ref{eq:nd4}.
\begin{eqnarray}
\dot{x}_i &=& x_i (r_i + \sum^{N}_{j=1} a_{ij} x_j), i = 1, ..., N,\label{eq:nd1}\\
X &=& [x_1, ..., x_{N}]^T,\\
  &=& [E_1, ..., E_{n_1}, V_1, ..., V_{n_2}, S_1, ..., S_{n_3}]^T,\\
N &=& n_1 + n_2 + n_3,
\label{eq:nd4}
\end{eqnarray}

\noindent where $X$ is an $N$-dimensional vector that consists of $N$ state variables including $n_1$ economic factors, $n_2$ environmental factors, and $n_3$ social factors.

\begin{figure}[h!]
	\centering
	\includegraphics[scale=0.35]{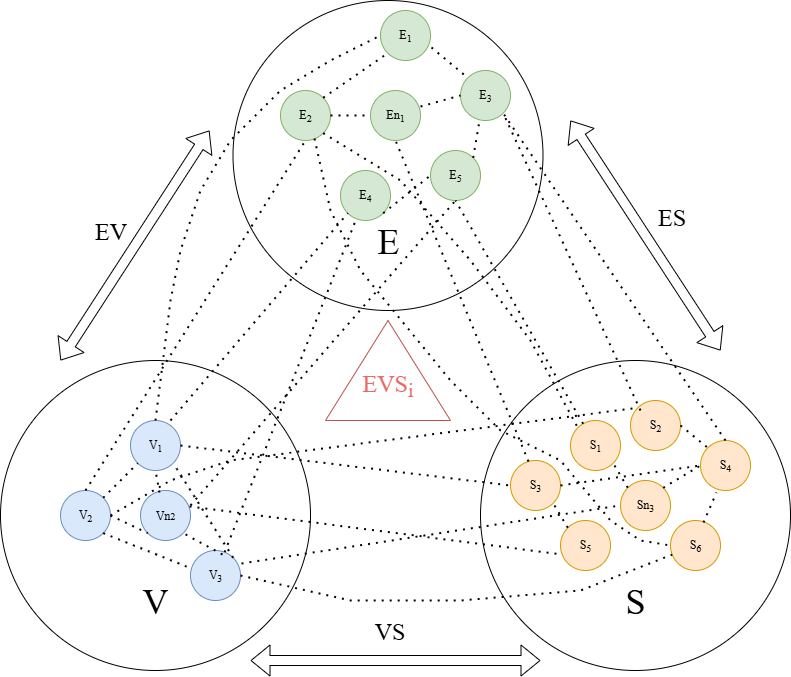}\\
	\caption{An illustration of the $N$-dimensional EVS model}
	\label{fig:EVS_n}
\end{figure}


The $N$-dimensional EVS system may exhibit substantially greater complexity, compared with the highly stylised 3D baseline EVS model discussed in the previous section. These contributing factors may exhibit random cooperative or competitive relationship amongst each other, in contrast to the relatively clearer relationship in the baseline setting. Furthermore, the full system can be decomposed into many interconnected subsystems, within which characterise the sustainability risks related to a particular issue. For example, one may take the most relevant $E_i$, $V_i$, $S_i$ to examine how the development of AI may lead to sustainability risks within the sub-EVS system of AI triangle, which is also connected with the full system. The subsystems and the full system can be aggregated and disaggregated with a great degree of flexibility for the sake of tackling a particular problem set of our interests, as well as tracking their interdependence with other subsystems and the full system, as illustrated in Fig. \ref{fig:EVS_n}. 

The full analysis of the $N$-dimensional EVS model is beyond the scope of this paper, which will be further investigated in future research. There are, however, many existing studies in the field of applied mathematics on high-dimensional Lotka-Volterra systems, which may be relevant to understanding the full EVS model in our context. \citet{may1972} demonstrates that for large linear systems with random coupling, the fixed points must jointly satisfy a fundamental inequality to ensure stability. May's inequality remains a cornerstone of theoretical ecology, inspiring decades of research on complexity management in networks. \citet{GOUZE1993} examines a generalised $N$-dimensional Lotka-Volterra (LV) model, which fundamentally transforms the analysis of global dynamics in the LV systems by reducing the problem to a matrix-algebraic framework. This approach rigorously establishes the universality of regular behaviour characterised by trajectories converging either to equilibria or diverging to infinity – across systems satisfying decomposition criteria. However, establishing the necessary and sufficient conditions for the existence of such decompositions remains an unresolved issue. \citet{akjouj2024complex} propose an analytical framework based on a stochastic matrix for large-scale Lotka-Volterra models, which extends \citet{may1972} and provides a rigorous mathematical foundation for understanding the stability of complex ecosystems. Future research may further look at the empirical estimation of the interactive matrix in the EVS system to understand the full scope of competitive or cooperative relationships among the $N$ variables. This can potentially be done by integrating a machine learning approach to overcome analytical difficulties and computational bottlenecks in high-dimensional systems.

Another potentially interesting issue associated with the full model is the possible existence of chaos in a higher-dimensional model when $N>3$. Numerous studies investigate the chaotic dynamics in Lotka-Volterra systems. \citet{Vano_etal_2006} found that chaos can exist in a Lotka-Volterra model when the system consists of at least four dimensions. \citet{christie2001chaos} rigorously demonstrates, via Melnikov's method, that periodic perturbations can induce Smale horseshoe chaos in Lotka-Volterra systems. This study establishes explicit parameter threshold conditions for chaotic behaviour and extends this framework to biomolecular self-organisation models, and establishes a novel paradigm for analysing complex dynamics in ecology and biophysics. 

Last but not least, the recent developments in dynamical mean field theory (DMFT) would also be relevant to the full analysis of the EVS system. DMFT is developed and applied to analyse dynamic systems consisting of a large number of interacting agents by reducing it to a one-dimensional master equation in terms of mean-field interaction. Its application in social sciences is popularised by \citet{WeidlichHaag1983}. More recently, \citet{Roy_2019} adopts DMFT to an $N$-dimensional Lotka-Volterra model, which is relevant in our context of an $N$-dimensional EVS system. The full mathematical properties of the generalised EVS system deserve further investigation in future research.

%
\section{Conclusion}
\label{Conclusion}
The extreme weather conditions resulting from climate change and other environmental catastrophes in recent years have served as a profound wake-up call for every global citizen to rethink the sustainability of our current economic paradigm. On a narrow level, some insurance companies start to realise that many of their properties are no longer insurable under the old actuarial framework due to climate change. On a broader level, the apocalyptic prediction of \citet{M1972} has started to gain renewed attention in recent years. Sustainability risks concern the well-being of every living being on the planet and demand a continuing commitment and global cooperation, in liaison with the UN's efforts and guided by the scientific community.

While global warming is an insidious process driven by accumulated fuel consumption over the past century, disruptive technologies emerging today may pose sustainability risks with much greater social and ecological impacts over a shorter time horizon. It is therefore vital to systematically monitor and manage sustainability risks at an early stage of the technological dissemination to prevent unforeseen social and ecological disasters. To this end, this paper proposes the EVS model: a generalised model of sustainability that characterises the interdependence between the economic ($E$), environmental ($V$), and social ($S$) factors in alignment with the sustainability definition of \citet{WCED1987}. The EVS model connects all the state variables in the three sub-dimensions through their competitive and cooperative relationships, and thus can be applied to understand the sustainability risks of disruptive technology in a systematic manner. The EVS model is inspired by earlier works of \citet{M1972}, \citet{M1992}, \citet{Herrington2021} and \citet{KM2011}. Our model has several advantages that can address the earlier criticisms of the \textit{World3} model. \textit{First}, the model has a better analytical tractability thanks to its simplistic mathematical structure based on a Lotka-Volterra setting. Despite its simplicity, the system can fairly characterise the competitive and cooperative relationship amongst the state variables of our interests. \textit{Second}, we adopt a Schumpetarian definition of economic development to better characterise the innovation-induced sustainability risks, rather than the relatively static assumption of technological development in the \textit{World3} model. \textit{Third}, the model has a high degree of flexibility, as it can be decomposed into multiple subdynamics to address specific sustainability risks within a particular sector, as well as to establish connections between the subsystem and other components, thereby understanding sustainability risks from a holistic perspective. The model can also shed light on the boundary between the private and public sectors in maintaining sustainable development. While the private sector is driven by Schumpeterian-type disruptive innovation and technological dissemination, sustainability risks can arise, which require the public sector to constantly monitor and manage its sustainability risks and ensure that the control parameters remain confined to the sustainability regime by enforcing better environmental regulation, encouraging innovation in the circular economy, and fostering social development that ultimately serves the economy and the environment.

The current version of our EVS model is still in its infancy, which can be extended and applied in numerous directions, both empirically and theoretically. While ESG can still serve as a means to promote corporate sustainability, EVS can be applied to evaluate sustainability risks associated with the existing and emerging technologies, particularly regarding their roles in exacerbating or mitigating climate risks, as well as other environmental and social risks, to prevent catastrophic consequences beyond the point of control. These extensions and applications will be further investigated in future research.

\newpage
\appendix
\section{The fixed point resilience for the ES and VS subdynamics}
\label{app_1}

\begin{condition}[Environmental resilience of the ES subdynamics]
\label{cond:ES}
When the ES subsystem approaches its fixed point \((E^*, S^*)\),
\begin{equation}
\frac{dV}{dt} \bigg|_{(E,S) \to (E^\star,S^\star)} > 0,
\end{equation}
\end{condition}

\begin{condition}[Economic  resilience of the VS subdynamics]
\label{cond:VS}
When the VS subsystem approaches its fixed point \((V^*, S^*)\),
\begin{equation}
\frac{dE}{dt} \bigg|_{(V,S) \to (V^\star,S^\star)} > 0.
\end{equation}
\end{condition}

To determine the environmental sustainability condition, we set $V = 0$. The system is reduced to an ES subdynamics:\\\\
\begin{equation}
\left\{
\begin{aligned}
\frac{dS}{dt} &= S(r_S 
+ a_{31}E) = 0, \\
\frac{dE}{dt} &= E(r_E + a_{13}S) = 0,
\end{aligned}
\right.
\end{equation}

\noindent which yields

\begin{equation}
\left\{
\begin{aligned}
S^* &= -\frac{r_{E}}{a_{13}}, \\
E^* &= -\frac{r_S}{a_{31}}.
\end{aligned}
\right.
\end{equation}

This fixed point may be unattainable since the sign of $a_{31}$ is undetermined. In case if $a_{31} > 0$, both $S^\star$ and $E^\star$ take positive signs. The environmental sustainability condition requires that:
\begin{equation}
\frac{dV}{dt} = V(r_V + a_{21} E + a_{23} S) > 0.
\end{equation}\\

Thus, we have

\begin{equation}
a_{31} (a_{13}r_V - a_{23}r_E) > a_{21}a_{13}r_S.
\end{equation}

Since, $r_V>0$, $r_E<0$, $a_{31} > 0$, $a_{13} > 0$, and $a_{23}>0$, it implies that $a_{13}r_V - a_{23}r_E > 0$. We have:
\begin{equation}
a_{31} > \frac{a_{21}a_{13}r_V}{a_{13}r_V - a_{23}r_E}. 
\label{eq:SE}
\end{equation}

Similarly, to determine the economic sustainability condition, we set $E = 0$. The VS subdynamics is set as follows:
\begin{equation}
\left\{
\begin{aligned}
\frac{\mathrm{d}V}{\mathrm{d}t} &= V(r_V + a_{23}S) = 0, \\
\frac{\mathrm{d}S}{\mathrm{d}t} &= S(r_S + a_{32}V) = 0,
\end{aligned}
\right.
\end{equation}

\begin{equation}
\left\{
\begin{aligned}
S^* &= -\frac{r_V}{a_{23}}, \\
V^* &= -\frac{r_S}{a_{32}}.
\end{aligned}
\right.
\end{equation}

This fixed point is unattainable since $r_V > 0$ and $a_{23} > 0$, which implies that $S^\star < 0$.

\newpage
\section{Numerical Simulation of the 3D Baseline EVS Model}
\label{appendix_NS}
\begin{figure}[h!]
\centering
    \caption{The EVS Dynamics when $a_{31} > 0$}
  \begin{subfigure}{9cm}
    \centering
    \includegraphics[width=1\linewidth]{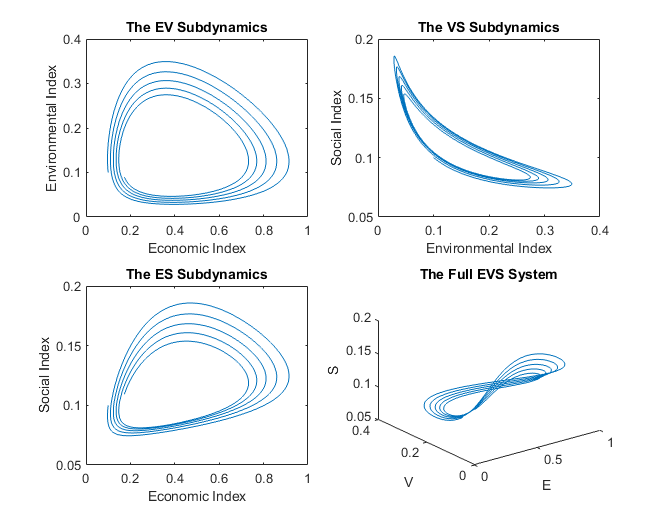}
    \caption{The EV, ES, VS Subdynamics and the EVS 3D Plot}
    \label{fig:EVS1}
  \end{subfigure}
  
  \begin{subfigure}{9cm}
        \centering
    \includegraphics[width=1\linewidth]{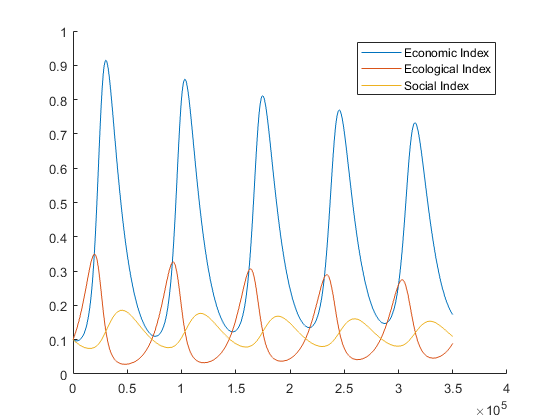}
    \caption{The EVS Individual Plots}
    \label{fig:EVS2}
  \end{subfigure}
  \label{fig:EVS}
\end{figure}
\newpage
\begin{figure}[h!]
\centering
    \caption{The EVS Dynamics when $a_{31} < 0$}
  \begin{subfigure}{9cm}
    \centering
    \includegraphics[width=1\linewidth]{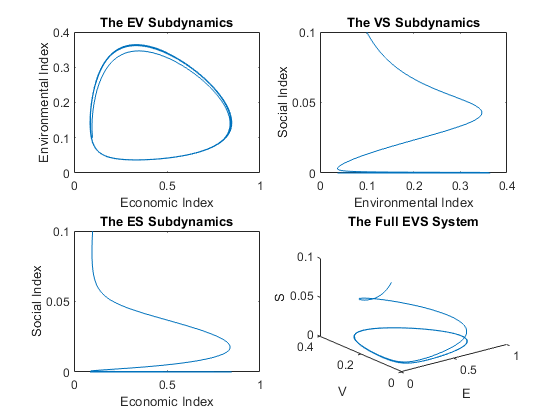}
    \caption{The EV, ES, VS Subdynamics and the EVS 3D Plot}
    \label{fig:EVS3}
  \end{subfigure}
  
  \begin{subfigure}{9cm}
        \centering
    \includegraphics[width=1\linewidth]{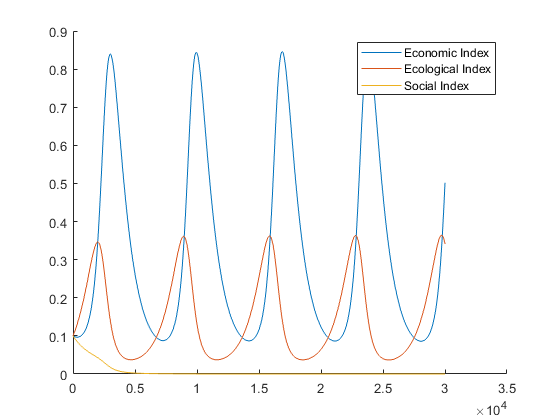}
    \caption{The EVS Individual Plots}
    \label{fig:EVS4}
  \end{subfigure}
  \label{fig:EVSS}
\end{figure}
\newpage
\begin{figure}[h!]
\centering
    \caption{Sensitivity of the EVS system to variations of $a_{12}$}
  \begin{subfigure}{9cm}
    \centering
    \includegraphics[width=1\linewidth]{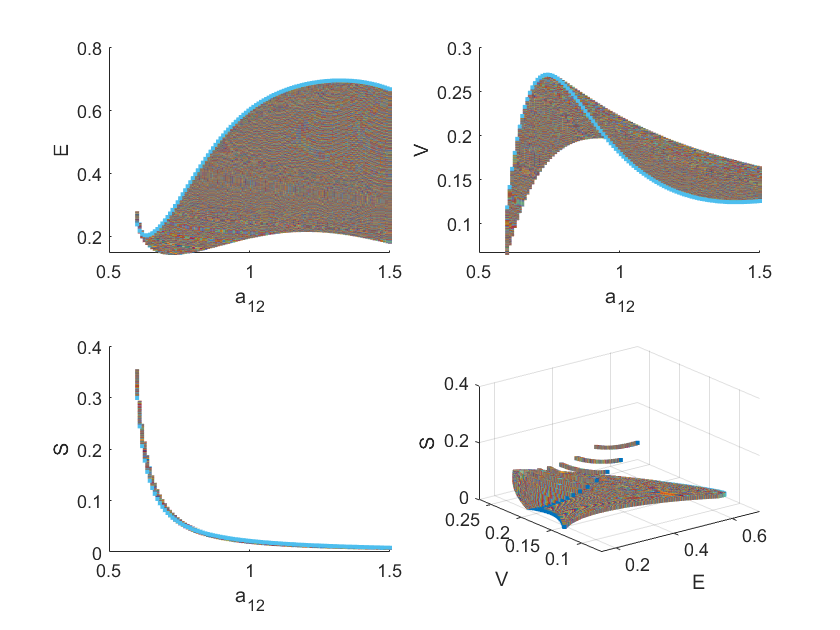}
    \caption{$a_{31} > 0$}
    \label{fig:EVS1_bif}
  \end{subfigure}
  
  \begin{subfigure}{9cm}
        \centering
    \includegraphics[width=1\linewidth]{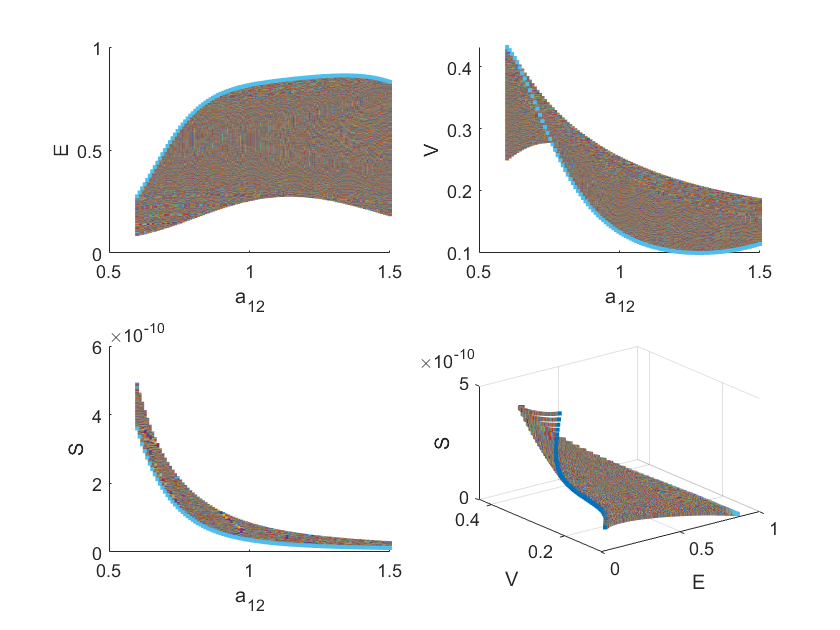}
    \caption{$a_{31} < 0$}
    \label{fig:EVS2_bif}
  \end{subfigure}
  \label{fig:EVS_bif}
\end{figure}
\newpage
\section*{Author Contributions}
\noindent The authors are listed in alphabetical order, and both authors have made equal contributions to this paper. This project is initiated by Tianhao Zhi, following his long-term research interests in sustainable development and nonlinear economic dynamics. Tianhao has contributed to the initial idea of applying a Lotka-Volterra model to sustainability risks, conducted a substantial part of the literature review, outlined the modelling structure, and wrote the first draft of this paper and the MATLAB code for numerical simulations. He presented the preliminary version of this paper at the \textit{International Day for Biodiversity 2024 cum Guangdong-Hong Kong-Macao Greater Bay Area Nature Education and Ecological Economic Forum} organised by Dr Siu-Tai Tsim, Associate Head of the Department of Life Sciences, Beijing Normal-Hong Kong Baptist University. At a later stage, Tianhao Zhi invited Yiren Wang to join the project after a period of regular discussion with each other on this topic. Yiren has made a substantial contribution by surveying the technical literature regarding recent developments in the Lotka-Volterra model, formulated the mathematical definition of sustainability, and provided the mathematical derivations and proofs of the 3D model. The remaining errors are on our own.

\section*{Acknowledgement}
\begin{itemize}
\item Our work was supported by the Guangdong Provincial Key Laboratory of IRADS (2022B1212010006).
\item The authors would like to thank Howard Huxter for introducing the two works of the \textit{Club of Rome} to Tianhao Zhi a few years ago: \textit{Limits to Growth} and \textit{Mankind at the Turning Point}. It serves as an inspiration for the Tianhao's continuing research on sustainable development.
\item The authors would like to thank participants at the \textit{International Day for Biodiversity 2024 cum Guangdong-Hong Kong-Macao Greater Bay Area Nature Education and Ecological Economic Forum} in Zhuhai, China on 18 May 2024, the \textit{2025 WEHIA (Winter) Workshop of Economics with Heterogeneous Interacting Agents} at XJ-Liverpool University in Suzhou, China (13 - 16 February 2025), and the \textit{2025 Risk Sciences Annual Conference} at the Nanyang Technological University in Singapore on 10 September 2025 for valuable feedback and suggestions.
\item The usual disclaimer applies.
\end{itemize}

\section*{Declaration}
\noindent  The authors have no competing interests to declare that are relevant to the content of this article.

\newpage
\bibliographystyle{elsarticle-harv}
\bibliography{WZ_2025}



	
	
	

\end{document}